# Spatial-temporal manipulations of visible nanosecond sub-pulse sequences in an actively Q-switched Pr:YLF laser

Shengbo Xu[†], Yunru Chen[†], Ran Xia[*], Changcheng Duan, Qingrui Zeng, Yu Xiao, Xiahui Tang, Gang Xu[*]

*Abstract*—Pulsed visible lasers either by Q-switching or mode locking have been attracting intense attentions both in solid-state laser and fiber laser. Here, we report on the simultaneous manipulation of reconfigurable sub-pulse sequences and customizable high-order vortex beams in an actively Q-switched visible laser. On the one hand, pulse sequences with up to 4 sub-pulses could be generated and fully controlled by means of an acoustic-optic modulator driven by an arbitrary waveform generator. Both pulse number and pulse intensity can be manipulated through the programmable step-signal, which is also theoretically simulated through the rate equations. On the other hand, assisted by the off-axis pumping technique and the astigmatic mode conversion, the laser cavity could emit high-quality vortex beams carrying Laguerre-Gaussian modes up to $30^{th}$ order. To the best of our knowledge, this is the most flexible active manipulations not only on the intensity distribution of the transverse modes but also on the temporal distribution of the pulse sequences in a visible laser. The versatile manipulating techniques in this work could be immediately implemented into all other solid-state lasers to obtain sub-pulse vortex beams, which may provide enhanced functionality and flexibility for a large range of laser systems.

## I. INTRODUCTION

LIGHT travels fast, giving rise to delicate and challenging management of photon propagation. Owing to the invention of laser in the 1960s', the manipulation techniques of optical field via functional photonic devices, either inside or outside the cavity could remarkably enable the improvement of their output performances [1], [2]. With this aim, fast-response modulation modules, e.g. the electric optic modulators (EOM) have been implemented into laser devices to harness the temporal (via intensity/phase modulators) or spatial (by spatial light modulators) manipulation of laser field [3], [4], [5]. Taking advantage of the programmable driving RF signals with high-repetition rates, one may tremendously broaden the laser tunability on their center wavelength, bandwidth, phase, coherence, polarization, longitudinal/transverse modes and so on [6]. The potential of laser has been excavating and extending with the development of the modulation devices, bringing forth the most powerful tool either in scientific research and applications.

In this framework, one of the most well-known scenarios is the active Q-switching [3], [6], where high-energy laser pulses could be emerged through the intracavity loss management with either the EOMs or the acoustic optic modulators (AOMs). For diode-pumped solid-state lasers, the later scheme is adequate to produce ns pulses with higher repetition rate (dozens of KHz) without requiring KV-level driving signal [2], [6], which is practical for a large range of applications such as laser drilling, cutting, laser medical treatment and laser induced breakdown spectroscopy (LIBS) [1], [2], [3]. [4], [5], [6]. Compared to a single giant Q-switched pulse, sub-pulse sequences with nanosecond or microsecond intervals are found to be more appropriate for aforementioned applications in recent years, which could be obtained either through selecting pulses from high-repetition-frequency lasers or by splitting pulses from low-repetition-frequency lasers [7]. Apart from those regeneration method, researches have demonstrated the emergence of Q-switched double pulses when the AOM is driven by stepped signals [8], [9], [10]. Such simple but practical scheme provides the versatile Q-switched pulses exhibiting more salient properties. Particularly, precisely design of the temporal interval of the sub-pulses (twins or triplets) to match the thermal phase transition of the material can remarkably boost the processing resolution, especially the depth-to-diameter ratio for drilling and cutting. [11], [12]. Furthermore, [13], [14], controllable Q-switched sub-pulses sequences are strongly desirable for high-resolution LIBS because the strict limitation of the delay between the leading pulse (material ionization) and the following pulse (reference for monitoring on plasmonic radiation) can be eliminated [15], [16], [17].

Apart from the aforementioned temporal distribution, the transverse modes providing additional degrees of freedom have been attracting growing research attention in the past three decades. Particularly, the vortices of light characterized by

Manuscript received March the $21^{st}$ 2024.
This work was supported by Key developing project of Hubei Province of China under the grant 2022BAA009；National Natural Science Foundation of China under the grant 62275097.

[†]These authors contributed equally to this work.

([*]Corresponding Author: rxia@hust.edu.cn, gang_xu@hust.edu.cn).
All the authors are with School of Optical and Electronic Information, Huazhong University of Science and Technology, Wuhan, China.

optical angular momentum (OAM) which could be converted from fundamental Gaussian beams through manipulating the wavefront phase in spatial domain have been showing the potential applications on optical telecommunications [23], [24], particle manipulations [25], [26], [27], micromachining [28], [29], quantum information [30] and so on. It is worth mentioning that the pulses vortex beams bearing the OAM with specific topological charges are of particular interest not only for the research on high-capacity optical communications, but also to create extreme conditions for light-matter interactions [44]. Especially, visible vortex pulses are reported to be achieved in passive Q-switching or mode-locking cavities such as visible [35], [36], [37], [38] or infrared [39], [40], [41], [42] ultrafast pulse trains with vortex beam shape.

However, pulsed vortex beams are rarely reported to be generated in active Q-switched or active mode-locked lasers, not to mention the manipulations on the temporal and spatial distribution of visible pulses. In this paper, we demonstrated a diode-pumped Q-switching Pr:YLF laser whose pulse sequences could be managed both in temporal and spatial domain. Utilizing the acoustic-optic modulator driven by the programmable step-signal, sub-pulse sequences with different pulse number and pulse intensity can be generated, which is also theoretically verified. Moreover, high order modes can be achieved by the off-axis pumping scheme including translation-based pumping and angle-based non-collinear pumping. Consequently, the reconfigurable sub-pulse sequences with customizable high-order OAM can be obtained with the help of extra-cavity astigmatic mode conversion. For the first time, we have fertilized the multi-step Q-switching technique and the off-axis pumping scheme to realize the output of 4 Q-switched sub-pulses in a single roundtrip with the Laguerre-Gaussian (LG) beam up to the 30th order. This method could be implemented to all other solid-state lasers to generate reconfigurable giant vortex laser pulses.

## II. NUMERICAL SIMULATION

The intra cavity dynamics of the photon number $\phi$, the population inversion number $N$ and the photon lifetime $\tau_c$ in an acoustic-optic Q-switching laser could be described with the following coupled differential equations [1]:

$$\begin{cases} \dfrac{\partial N}{\partial t} = R_p - N(\gamma\sigma\phi - \dfrac{1}{\tau_f}) \\ \dfrac{\partial \phi}{\partial t} = V\gamma\sigma c\phi N - \dfrac{\phi}{\tau_c} \\ \tau_c = \dfrac{\tau_r}{\delta_{AOM} - \ln R_0 + L_i + T_{OC}} \end{cases} \quad (1).$$

Here, $R_p$ is the pump rate and $\gamma$ is the constant factor about the decrease of the population inversion (equal to 1 for this case). $\tau_f$ and $t_r$ are the fluorescence lifetime and the cavity roundtrip time of the upper energy level, respectively. $V$ and $\sigma$ refer to the volume of the gain area and the emission cross section of the Pr:YLF crystal. And $\delta_{AOM}$, $lnR_0$, $L_i$ and $T_{OC}$ are cavity losses caused by the AOM, mirror coupling, diffraction and the output coupler, respectively. Thus, the generation of Q-switching pulse sequences [2],[3] [49] with the variation of intracavity loss q could be revealed as shown in Fig. 1.

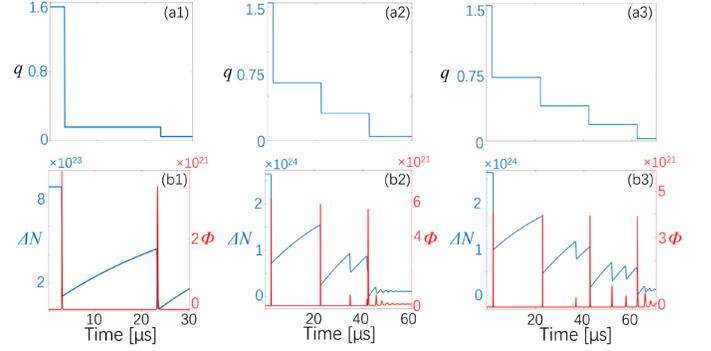

**Fig. 1.** Numerical simulation of the sub-pulse generation via multi-step Q-switching. (a1-a3): the step-like variation of intra-cavity Loss $q$ with different steps from 2 to 4; (b1-b3): the corresponding results of population inversion (blue line) and sub-pulse sequence (red line), respectively.

Theoretically, the pulses would arise when the inversed populations were released at each falling edge of the controlled loss $q$, resulting in the Q-switched sub-pulse sequence including two, three, four pulses as shown in Fig. 1(b1-b3). Particularly, despite of the invariable value between two falling edges of $q$, the inversed population keeps rising during 20 μs delay due to the constant pumping. The peak intensity of a laser pulse is determined by the certain amount of released population associated with the total cavity loss $q$. In order to generate sub-pulses with the same peak intensity simultaneously, the former sub-pulse needs to consume more inversed population as shown in Fig. 1(b1), resulting from the careful design of the loss control. This phenomenon is also manifested in the cases of three and four sub-pulses simulations showed in Fig.1(b2) and (b3). However, several tiny spikes appear during the delay between sub-pulses when the pulse number getting larger, demonstrated both in Fig.1(b2) and (b3). A novel numerical simulation about the variation of the spiking interval during the start of the relaxation oscillation [43] reveals the theory about the tiny spikes around sub-pulses. Relaxation oscillation (spiking) is more likely occurred in a high threshold population inversion laser system as sub-pulses lasers. The spikes which are arose by relaxation oscillation have a certain pattern that varied spiking period and the spiking intensity decays. Population inversion, cavity loss and pump intensity jointly decide the temporal interval between each spike. As shown in Fig. 2, the cavity loss $q$ remains minimum after the generation of the single pulse and the spikes appear after the giant Q-switching pulse with a varied spike period as mentioned above. Those spikes around sub-pulses is similar to this result but the subsequent variation of the total loss $q$ prevent the generation of following spikes and the continuous wave. Numerical simulation results in Fig. 2 (b) provides the reference for the design of the AOM driving signal and implied the possibility of spikes inhibition by modulation of the pumping. It can be foreseen that enhance the pump power is able to boost sub-pulses increasement as the maximum population inversion before each sub-pulses sequence generation influences the maximum number of sub-pulses.

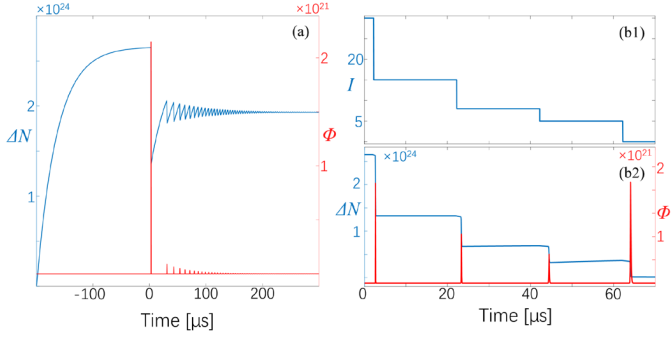

**Fig. 2.** Numerical simulation on spikes caused by relaxation oscillation in different pump intensity( a): The following spikes after a giant Q-switching pulse generation while the pump intensity is a constant. (b): Inhibition of the spikes in four sup-pulses via modulation of the pumping, $I_{pump}$ refers to the pump intensity.

## III. EXPERIMENTAL SETUP

Figure 3 (a) depicts the schematic illustration of the experimental setup. The gain medium is a *c*-cut Pr:YLF cuboid crystal (3×3×15 mm$^3$) with anti-reflective coating for the wavelength range of 400-700 nm, which is end-pumped by a home-made external cavity diode laser (ECDL) array with the output power up to ~ 40 W. And the pumping wavelength could be reliably locked at 444±0.1 nm with the full linewidth at half maximum less than 0.8 nm by a feedback blazed grating. This extra-cavity design of the pump is favorable to match the narrow absorption peak of the Pr$^{3+}$ [45], [46], [47], boosting the pumping efficiency and avoiding the heat damage and the related thermal lens effect [48]. Moreover, the crystal is wrapped with indium foil and mounted in a copper heat sink with water cooling at 12 ℃. The gain crystal is placed in an 80 mm-long plano-concave cavity with the curvature radius of 50 mm. Here the input plain mirror (IM) and the output concave mirror (OC) are processed with the dichroic coating (HT@444 nm, HR@500-750 nm). The later has the coating with 3% transmittance at the 500-750 nm to serve as an output coupler. Both IM and OC are placed in standard mirror mounts with three adjusters to achieve off-axis pumping. The output is split by a 1:9 beam splitter (BS). 1% portion is detected by a fast photodetector () and the oscilloscope (Siglent, SDS2352X-E) to measure the pulse intensity profile. The rest is sent to a pair of cylindrical lens (CL1 and CL2) for the purpose of astigmatic transformation, so that the Hermit-Gaussian (HG) modes can be transferred to LG modes [39], [49]. In order to gain more physical characteristics of the vortex beams, we build a Michelson interferometer and the self-interference patterns are recorded with a CMOS camera (Femto Easy beam analyzer, μ-BP7.6).

With appropriate optimization of the cavity, the CW laser at 639 nm can be achieved with maximum output power of 1.05 W under 6W pump. As shown in Fig. 3.(b), the pumping threshold $P_{th}$ was 2.1W and the slope efficiency was about 40%. Furthermore, we inserted an AOM (Waveguard laser, WGQS-080-1.2-CQC-1064-AF) inside the cavity to realize the acoustic optic Q-switching. As shown in Fig. 3(c), the step-like driving RF signal generated by an arbitrary function generator (AFG) (Siglent, SDG 2122X) lead to the emergence of Q-switched pulse with the duration of about ▮s, which is determined by the cavity length because the inversed population was completely released. Specially, a delay of ~ 1.3 μs between the pulse and the falling edge of the RF signal resulting from the μs-scale response time of the AOM was observed. The 10-KHz Q-switched pulse train shown in Fig. 1(d) could be obtained by a ▮Hz RF signal.

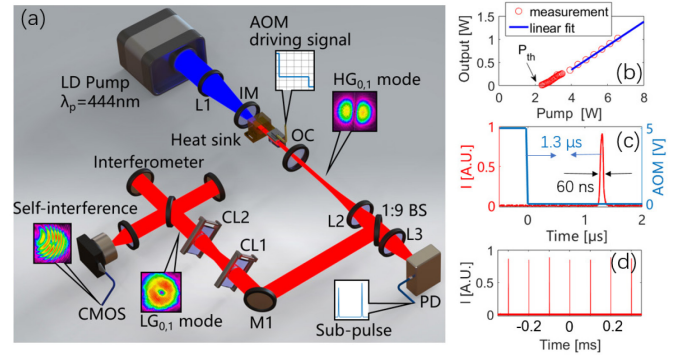

**Fig. 3.** (a) Experimental setup. LD: laser diode array; L1: lens with a focal length of 75 mm; IM: input plain mirror; OC: output concave mirror with a radius of curvature of -100 mm; BS: beam splitter; PD: photon diode; CL1 and CL2: cylindrical lens. (b) Dependence of output power and the pump power. The dark arrow indicates the pump threshold of $P_{th}$ ~ 2.1 W. Red circles are measured input/output power and the solid blue line is the corresponding linear fit. (c) Single-step AOM driving signal (blue curve) and the typical pulse intensity profile (red curve). (d) Q-switched pulse train [zoom out of (c)].

## IV. RESULTS AND ANALYSIS

The simultaneous manipulation of the temporal separation and pulse intensity, as well as the transverse modes of the Q-switched pulse sequence basically could be realized by combining the multi-step modulation and off-axis pumping. Particularly, direct generation of the vortex beam with high-order LG modes could also be obtained by the angle-based non-collinear pumping scheme. Compared with the conventional pulsed vortex beams obtained via the passive mode locking or Q-switching [35], [36], [37], [38], [39], [40], [41], [42], the pulse intensity and temporal interval can be precisely controlled through the careful design and editing of the driving signals on the AOM.

*A. Manipulations of multi-step Q-switched sub-pulses*

With this configuration, the Q-switched laser can exhibit fascinate controllable features with the editable driving signal on the AOM. Fig. 5 depicts the generation of 3 and 4 sub-pulses and the manipulations on pulse intensities. As shown in Fig. 4(a), with the AFG, we have designed the RF signal ($V_{PP} = 5V$) with 3 steps and the distance of 20 μs per each. Importantly, the pulse delay could be freely and individually controlled by varying the step distance of the driving signal. Moreover, the intensity of the pulses could be individually adjusted by modifying the relative loss drop and the interval of each step, we can arbitrarily control the delay and the intensity distribution of the sub-pulse sequences. With this idea, Fig. 4(b) demonstrates five representative cases by varying the driving signal, e.g. the increasing, the decreasing or equally-distributed sequences.

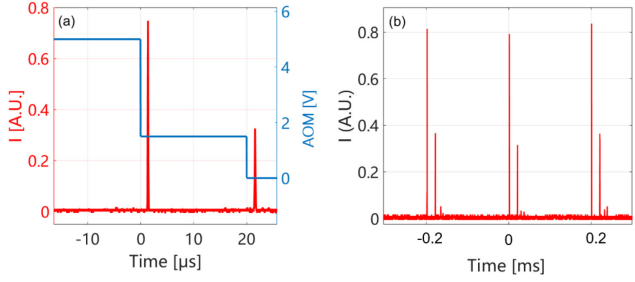

**Fig. 4.** Experimental results on the generation of two sub-pulses sequence. (a) The double-step driving signal on the AOM (blue curve) and the corresponding dual laser pulses. (b) Zoom-out of the dual-pulse train.

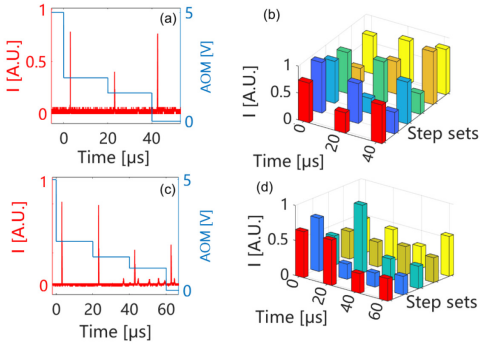

**Fig. 5.** Multiple sub-pulses generation and the pulse intensity manipulation. (a) Triple-step driving signal on the AOM (blue curve) and the output Q-switched pulses (red curve) with middle sagging sequence. (b) 5 typical pulses intensity distribution [high-low-high (red), decreasing (blue), high-low-low (cyan), increasing (orange) and equally distributed (yellow)] by varying the step drop of the driving signal. (c-d) 4 sub-pulses generation with the quadruplet step driving signal.

Similarly, we also performed the experiment to generate 4 sub-pulses by designing the quadruplet step driving signal on the AOM [see Fig. 5(c)&(d)]. Besides modifying the relative decline of step, the intensity of each pulse could be finely adjusted by turning the azimuthal angle of the AOM crystal. AOM is a b ragg diffraction device based on the refractive index grating arisen by ultrasonic wave in the quartz crystal, the incident angle determined the power distribution of both 0 and ±1 diffraction order. This may lead to the variation of intracavity losses, thus influencing on the intensity distribution of the pulse emission. We should also emphasize that when multi-step driving signal is added on the AOM, the sub-pulses could be emerged as expected, but they may be accompanied with the arise of the disordered small peaks around the trailing pulse. With the increasing of the driving steps, the noise spikes are more noticeable due to the parasitic oscillations inside the cavity.

*B. Manipulations on pulsed vortex laser beams*

Based on the sub-pulse Q-switching laser we focus on the simultaneous generation and the manipulations on the transverse modes to obtain the reconfigurable pulsed vortex beams. In optical resonators, the phase and intensity distribution of electromagnetic field may reproduce themselves upon repeated reflections between the mirrors [2]. So transverse mode beams, known as $TEM_{(m,n)}$ are all coaxial overlapped and occupy different volume in gain media. The schematic of off-axis pump are show in Fig. 6 (a) and (b), when an off-axis displacement was introduced between pump beam axis and cavity's optical axis, the fundamental mode $TEM_{(0,0)}$ fades out in mode competition, meanwhile oscillation shift to the high-order modes respectively. In the experiment setup shown in Fig. 1(a), the pupil of the laser system is the end face of the crystal. So with the diffraction of this square hole, the HG modes are more possibly to oscillate. While with the increases of tilt angle of OC, the beam profile changed from the standard $HG_{(1,1)}$ mode [Fig. 6 (c1)] to the $HG_{(1,2)}$ mode [Fig. 6 (c2)] and $HG_{(1,3)}$ mode [Fig. 6 (c3)]. Moreover, fine tuning the transversal displacement of the gain crystal and the its azimuthal angle may stimulate the oscillation of the LG modes, evident in Fig. 6 (d1), which demonstrates the direct generation of the $LG_{(0,1)}$ mode. To identify beam modes, Fig. 6 (d2) demonstrates the corresponding self-interference pattern which bears a clear furcation around the center with the clockwise rotation.

Fig. 6(d) illustrate a significant situation that the $LG_{(0,1)}$ mode could be directly generated in our laser cavity. However, the direct stimulation of higher-order LG modes inside the cavity could be more challenging due to the sensitivity of the loss and gain variation. In this work, we successfully obtain arbitrary order of stable LG mods by the transformation from the HG modes associated with the astigmatism mode converter (AMC) outside the cavity [50]. Transverse modes can be considered as light intensity distribution on the plane perpendicular to the ray direction. Assuming the $z$ axis corresponds to the direction of light travels, the amplitude of the HG mode (horizontal distribution direction 45° rotated) and the LG mode have the definitions by following equations:

$$\begin{cases} u^{LG}_{m,n}(x,y,z) = \sum_{k=0}^{N} i^k b(m,n,k) u^{HG}_{N-k,k}(x,y,z) \\ u^{LG}_{m,n}(\frac{x+y}{\sqrt{2}}, \frac{x-y}{\sqrt{2}}, z) = \sum_{k=0}^{N} b(m,n,k) u^{HG}_{N-k,k}(x,y,z) \end{cases} \quad (2).$$

Notice that all modes of LG modes and HG modes could be expanded into polynomials formed by HG modes in adjacent order HG modes. Here, the $b(m.n.k)$ is the polynomial coefficients, decided by following formula:

$$b(m,n,k) = [\frac{(N-k)!k!}{2^N m!n!}]^{\frac{1}{2}} \times \frac{1}{k!} \frac{d^k}{dt^k} [(1-t)^m (1+t)^n]_{t=0} \quad (3).$$

Those two polynomials figure out that the only difference between LG modes and HG modes is the imaginary unit, $i$, which refers to the phase shift of $\pi/2$. With this idea, we may convert the HG modes to the LG modes by implementing a pair of identical CL. In our experiment, the generatrix direction of both CL are fine tuned to exactly 45° rotated relative to the incident HG mode beam and placed just at a distance of $f/\sqrt{2}$ to induce the essential Gouy phase in specific directions [50]. With this simple but efficient AMC, we successfully obtained each order LG mode beam up to the 30[th] order converted from the corresponding HG modes. Fig. 6 lists four typical modes and transformed results from 1[st] order to 4[th] order and the 30[th] order LG mode. The topological charge number of each converting mode can be easily deduced by the original order of the HG mode. To gain more physical insights of the obtained LG modes, we recorded the self-interference patterns which are shown in Fig. 7 (c), and the clear furcation attribution in the figures refers to the order number. We should mention that with the order increases, the donut hole of the converted LG modes became larger, and the fringe of self-interference patterns are dencer, which is consistent with the natural law of the LG modes.

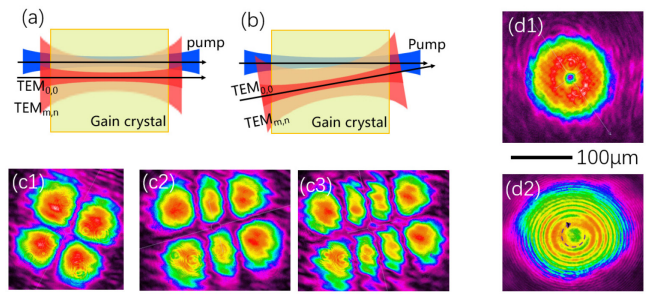

**Fig. 6.** Schematic of off-axis pump and the example of output transverse modes. (a) Off-axis pump caused by spatial translation of the crystal or the output coupling mirror; (b) Off-axis pump caused by spatial rotation of crystal; (c) Direct generation of high-order HG modes in the cavity: (c1) $HG_{(1,1)}$; (c2) $HG_{(2,1)}$; (c3) $HG_{(3,1)}$; (d1) Direct generated $LG_{(0,1)}$ mode and the corresponding self-interference spiral patterns (d2) via the Michelson interferometer [shown in Fig. 1(a)]. Dashed purple arrow in (d2) indicates the chirality of the angular momentum.

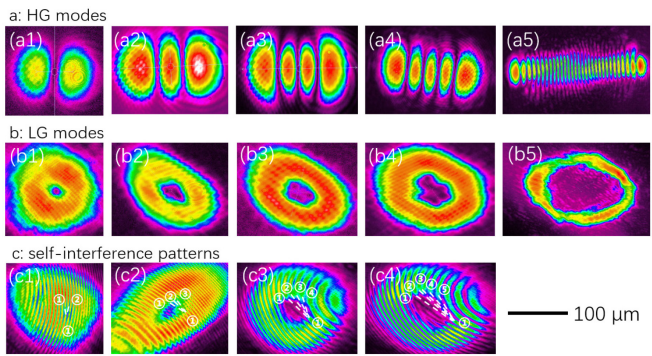

**Fig. 7.** Beam profiles of the sub-pulses with $HG_{(n,0)}$ modes patterns and their convert results. (a) original HG modes direct generated from Q-switching resonator, from $HG_{(1,0)}$ to $HG_{(30,0)}$, respectively; (b) LG modes converted from the HG modes above, respectively; (c) self-interference patterns, showing the topological charge of each LG modes.

## V. DISCUSSION AND CONCLUSIONS

In this work, we presented an effective and practical scheme aimed at the simultaneous generation of reconfigurable sub-pulse sequences and customizable high-order vortex beams in an actively Q-switched visible laser. By means of an AFG-driven AOM, we obtained 4 sub-pulses, whose temporal interval are fully controllable by judicially designing the driving step signals. By varying the step height, the pulse energy distribution between sub-pulses can be easily manipulated as desired. During the laser operation, we noticed that with step amount increasing, the noise-like spike became more obvious and generally gathered around the final sub-pulse. This phenomenon due to the relaxation oscillation [43], the spike will appear as a certain pattern that varied spiking period and the spiking intensity decays. The spike temporal interval associated with cavity loss and pump intensity. Predictably, the relevant spikes induced by the relaxation oscillation could be mitigated by modulating the pump power.

From the controllable Q-switching sub-pulses, manipulations on the transverse modes are performed to obtain the reconfigurable pulsed vortex beams. In this way, the laser cavity could emerge high-quality vortex beams with continuously varied HG modes up to the 30[th] order, and by means of an AMC which just combined with a CL pair, all HG modes can convert to corresponding LG modes. If required, more sub-pulse (>4) and higher-order (>30) vortex beams could be obtained by increasing the pump power. As this technique is based on very fundamental theories of laser cavity, it is capable to deploy to any active Q-switching lasers and obtain the giant pulsed vortex beams. From one hand, these practical manipulation techniques could enable various functionalities of pulsed lasers on manufacturing, LIBS, single molecular manipulation and so on. On the other hand, the unveiled tunability of the laser transverse modes, as well as the pulse timing in this work might provide more opportunities for the AI-based optimization to obtain the desired operation [52] [53]. Relevant algorithms developments and feedback-loop implementations require further investigations.

## REFERENCES

[1] H. Eichler, J. Eichler and O. Lux, Lasers, "Basics, Advances and Applications", *Springer Serises in Optical Sciences*, pp. 315-378, 2018.
[2] W. Koechner, "Solid-state laser engineering", *Springer*, pp. 452-559, 2002.
[3] C. Guo, S. Singh, "Handbook of laser technology and applications", *Taylor and Francis Group* (2021).
[4] C. Maurer et al., "What spatial light modulators can do for optical microscopy", *Laser Photonics Rev*, vol. 5, no. 1, pp. 81-101, 2011.
[5] B. Sun et al., "Four-dimensional light shaping: manipulating ultrafast spatiotemporal foci in space and time", *Light Sci. Appl.*, vol. 6, no. 1, pp. 293-299, 2017.
[6] U. Keller, "Ultrafast lasers", *Springer, pp.* 373-417, 2021.
[7] M. Hu et al., "Nanosecond double pulse fiber laser with arbitrary sub-pulse combined based on a spectral beam combining system", *Opt. Laser Technol.*, vol. 90, pp. 22-26, 2017.
[8] Y. Li et al., "Study of an acoustic-optic Q-swiched double pulse output Pr:YLF all solid-state laser", *Laser Phys.*, vol. 30, no. 12, pp. 125002.1-6, 2020.
[9] Z. Yang et al., "Blue LD-pumped electro-optically Q-switched Pr:YLF visible laser with kilowatt-level peak power", *Opt. Laser Technol.*, vol. 148, pp. 107711.1-6, 2022.
[10] J. Mao et al., "Three-nanosecond-equal interval sub-pulse Nd:YAG laser with multi-step active Q-switching", *Chin. Opt. Lett.*, vol. 19, no. 7, pp. 43-47, 2021.
[11] A. Asratyan et al., "Laser processing with specially designed laser beam", *Appl. Phys. A*, vol. 122, no. 4, pp. 434-439, 2016.
[12] J. Zhang et al., "Femtosecond laser double pulses nanofabrication on silicon", IOP Conf. Series: Material Science and Engineering, vol. 565, pp. 147-152, 2019.
[13] X. Wang et al., "Laser drilling if stainless steel with nanosecond double-pulse", *Opt. Laser Technol.*, vol. 41, pp. 148-153, 2009.
[14] W. Liu et al., "Double-pulse laser micro sintering: Experimental study and mechanism analysis aided by in-situ time-resolved tempreture measurements", *J. Manuf. Process*, vol. 69, no. 34, pp. 191-203, 2021.

[15] R. S. Harmon, "Laser induced breakdown spectroscopy", *Treatise on Geochemustry*, vol. 5, pp. 245-272, 2014.
[16] S. Legnaioli et al., "Industrial applications of laser-induced breakdown spectroscopy: a review", *Anal. Methods-UK*, vol. 12, no. 8, pp. 1014-1029, 2020.
[17] J. Skruibis et al., "Multiple-pulse laser-indued breakdown spectroscopy for monitoring the femtosecond laser micromachining process of glass", *Opt. Laser Technol.*, vol. 111, pp. 295-302, 2019.
[18] D. Stratis-Cullum et al., "Effect of pulse delay time on a pre-ablation dual-pulse LIBS plasma", *J. Spectro.*, vol. 55, no. 10, pp. 1297-1303, 2001.
[19] H. Sun et al., "Visible-Wavelength All-Fiber Mode-Locked Vortex Laser", *J. Lightwave Technol.*, vol. 40, no. 1, pp. 191-195, 2022.
[20] P. Coullet et al., "Optical vortices", *Opt. Commun.*, vol. 73, no. 5, pp. 403-408, 1989.
[21] Y. Shen et al., "Optical vortices 30 years on: OAM manipulation from topological charge to multiple singularities", *Light Sci. Appl.*, vol. 8, no. 5, pp. 697-725, 2019.
[22] C. Wan et al., "Optical spatiotemporal vortices", *eLight*, vol. 3, no .1, pp. 1-13, 2023.
[23] S. Franke-Arnold, L. Allen, M. Padgett, "Advances in optical angular momentum", *Laser Photonics Rev.*, vol. 2, no. 4, pp. 299-313, 2008.
[24] A. Willner et al., "Optical communications using orbital angular momentum beams", *Adv. Opt. Photonics*, vol. 7, no. 1, pp. 66-106, 2015.
[25] J. Wang, "Advances in communications using optical vortices", *Photonics Res.*, vol. 4, no. 5, pp. 1-15, 2016.
[26] M. Padgett, R. Bowman, "Tweezers with a twist", *Nat. Photonics*, vol. 5, no. 6, pp. 343-348, 2011.
[27] L. Paterson et al., "Controlled rotation of optically trapped microscopic particles", *Science*, vol. 305, no. 55188, pp. 912-914, 2001.
[28] C. Hnatovsky et al., "Material processing with a tightly focused femtosecond laser vortex pulse", *Opt. Lett.*, vol. 35, no. 20, pp. 3417-3419, 2010.
[29] K. Anoop et al., "Femtosecond laser surface structuring of silicon using optical vortex beams generated by a q-plate", *Appl. Phys. Lett.*, vol. 104, no. 24, pp. 241604.1-5, 2014.
[30] A. Mair et al., "Entanglement of the orbital angular momentum states of photons", *Nature*, vol. 412, no. 6844, pp. 313-316, 2001.
[31] V. Kotlyar et al., "Generation of phase singularity through diffracting a plane or Gaussian beam by a spiral phase plate", *J. Opt. Soc. Am. A*, vol. 22, no. 5, pp. 894-861, 2005.
[32] N. Matsumoto et al., "Generation of high-quality higher-order Laguerre-Gaussian beams using liquid-crystal-on-silicon spatial light modulators", *J. Opt. Soc. Am. A*, vol. 25, no. 7, pp. 1642-1651, 2008.
[33] Y. Zhang et al., "Tunable vortex beams generation in visible band via $Pr^{3+}$:YLF laser with a spot defect", *Appl. Phys. Lett.*, vol. 123, no. 25, pp. 25117.1-25117.6, 2023.
[34] J. Hao et al., "High-power vortex beams generated from a Yb-YAG thin-disk laser with spot-defect mirrors", *Opt. Laser Technol.*, vol. 169, pp. 110112-110117, 2024.
[35] N. Li et al., "High-order vortex generation from CW and passively Q-switched Pr:YLF visible lasers", *IEEE Photonic Tech. L.*, vol. 31, no. 17, pp. 1457-1460, 2019.
[36] N. Li et al., "Direct generation of an ultrafast vortex beam in a CVD-graphne-based passively mode-locked $Pr:YLF_4$ visible laser", *Photonics Res.*, vol. 7, no. 11, pp. 33-37, 2019.
[37] Q. Tian et al., "Direct generation of orthogonally polarized dual-wavelength continuous-wave and passively Q-switched vortex beam in diode-pumped Pr:YLF lasers", *Opt. Lett.*, vol. 44, no. 22, pp. 5586-5589, 2019.
[38] J. Zou et al., "Green-red pulsed vortex-beam oscillations in all-fiber lasers with visible-resonance fold nanorods", *Nanoscale*, vol. 11, no. 34, pp. 15991-16000, 2019.
[39] H. Liu et al., "High-order femtosecond vortives up to the 30th order generated from a powerful mode-locked Hermite-Gaussian laser", *Light Sci. Appl.*, vol. 12, no. 10, pp. 2084-2096, 2023.
[40] K. Yamane et al., "Ultrashort optical-vortex pulse generation in few-cycle regime", *Opt. Express*, vol. 20, no. 17, pp. 18986-18993, 2012.
[41] C. Lee et al., "Generation of higher order vortex beams from a $YVO_4$/Nd:$YVO_4$ self-Raman laser via off-axis pumping with mode converter", *IEEE J. Sel. Top. Quant.* vol. 21, no.1, pp 318-222, 2015.
[42] D. Lin et al., "The generaion of femtosecond optical vortex beams with megawatt powers directly from a fiber based", *Nanophotonics*, vol. 11, no. 4, pp. 847-854, 2022.
[43] S. Yoichi et al., "Initial Behavior of the Relaxation Oscillation at Zero-Phonon Line of Yb Gain Media", *CELO:2016*,
[44] Y. Fang et al., "Probing the orbital angular momentum of intense vortex pulses with strong-field ionization", *Light Sci. Appl.*, vol. 11, no. 2, pp. 276-286, 2022.
[45] B. Qu et al., "Broad-band tunable CW laser operation of $Pr^{3+}$:$LiYF_4$ around 900 nm", *Opt. Lett.*, vol. 40, no. 13, pp. 3035-3056, 2015.
[46] A. Richter et al., "Diode pumping of a continuous-wave $Pr^{3+}$ -doped $LiYF_4$ laser", *Opt. Lett.*, vol. 29, no. 22, pp. 2638-2640, 2004.
[47] X. Lin et al., "Diode-pumped wavelength-switchable visible Pr:YLF laser and vortex laser around 670 nm", *Opto-Electro. Adv.*, vol. 4, no. 4, pp, 27-34, 2021.
[48] P. Wang et al., "Thermal effect analysis on cuboid Pr:YLF crystals pumped by blue laser diodes", *Appl. Opt.*, vol. 62, no. 18, pp. 4797-4804, 2023.
[49] J. Kojou et al., "InGaN diode pumped actively Q-switched intracavity frequency doubling $Pr:YLF_4$ 261 nm laser", *Appl. Opt.*, vol. 53, no. 10, pp. 2030-2036, 2014.
[50] M. Beijersbergen et al., "Astigmatic laser mode converters and transfer of orbital angular momentum", *Opt. Commun.*, vol. 96, pp. 123-132, 1993.
[51] A. N. Chester, "Gain Thresholds for Diffuse Parasitic Laser Modes", *Appl. Optics*, Vol. 12, no. 9, pp. 2139-2146, 1973
[52] A. Naderi et al., "A review on applications of artificial intelligence in modeling and optimization of laser beam machining", *Opt. Laser Technol.*, vol. 135, pp. 106721, 2021.
[53] Q. Ma and H. Yu, "Artificial intelligence-enabled mode-locked fiber laser: a review", *Nanomanuf. Metrol.*, vol. 6, no. 36 (2023)